\documentclass[journal]{IEEEtran}
\usepackage{ifpdf}
\usepackage{graphicx}

\usepackage{cite}

\begin{document}
%
\title{Development Of Nanowire Devices With Quantum Functionalities}

\author{Michael~Stuiber,~Laurens~H.~Willems~van~Beveren,~Brett~C.~Johnson,~Walter~M.~Weber,
Andr\'{e}~Heinzig,~J\"{u}rgen~Beister,~David~N.~Jamieson,~and~Jeffrey~C.~McCallum
      \thanks{M. Stuiber, B. C. Johnson, D. N. Jamieson and J. C. McCallum are with the Centre of Excellence for Quantum Computation and Communication Technology, CQC2T. 
\newline
 E-mail: mstuiber@student.unimelb.edu.au.}
 \thanks{M. Stuiber, L. H. Willems van Beveren, B. C. Johnson, D. N. Jamieson and J. C. McCallum are with the School of Physics Department, University of Melbourne, Parkville, VIC $3010$, Australia.}
 \thanks{W. M. Weber, A. Heinzig and J. Beister are with Namlab gGmbH, N\"{o}thnitzer Str. $64$, $01187$ Dresden, Germany.}
}
\maketitle



%
\IEEEpeerreviewmaketitle

\section{Introduction}
%
%
%
%
\IEEEPARstart{S}{}ilicon has dominated the microelectronics industry for the last $50$ years. With its zero nuclear spin isotope ($^{28}$Si) and low spin orbit coupling, it is believed that silicon can become an excellent host material for an entirely new generation of devices that operate under the laws of quantum mechanics \cite{FZwanenburg}. Semiconductor nanowires however, offer huge potential as the next building blocks of nano-devices due to their one-dimensional structure and properties \cite{CThelander}. \\
We describe a fabrication process to prepare doped vapor-liquid-solid (VLS) grown silicon nanowire samples in a $2$- and $4$-terminal measurement setup for electrical characterisation.

\section{Fabrication Process}

\begin{figure}[htbp]
\centering 
\includegraphics[width=3.5in]{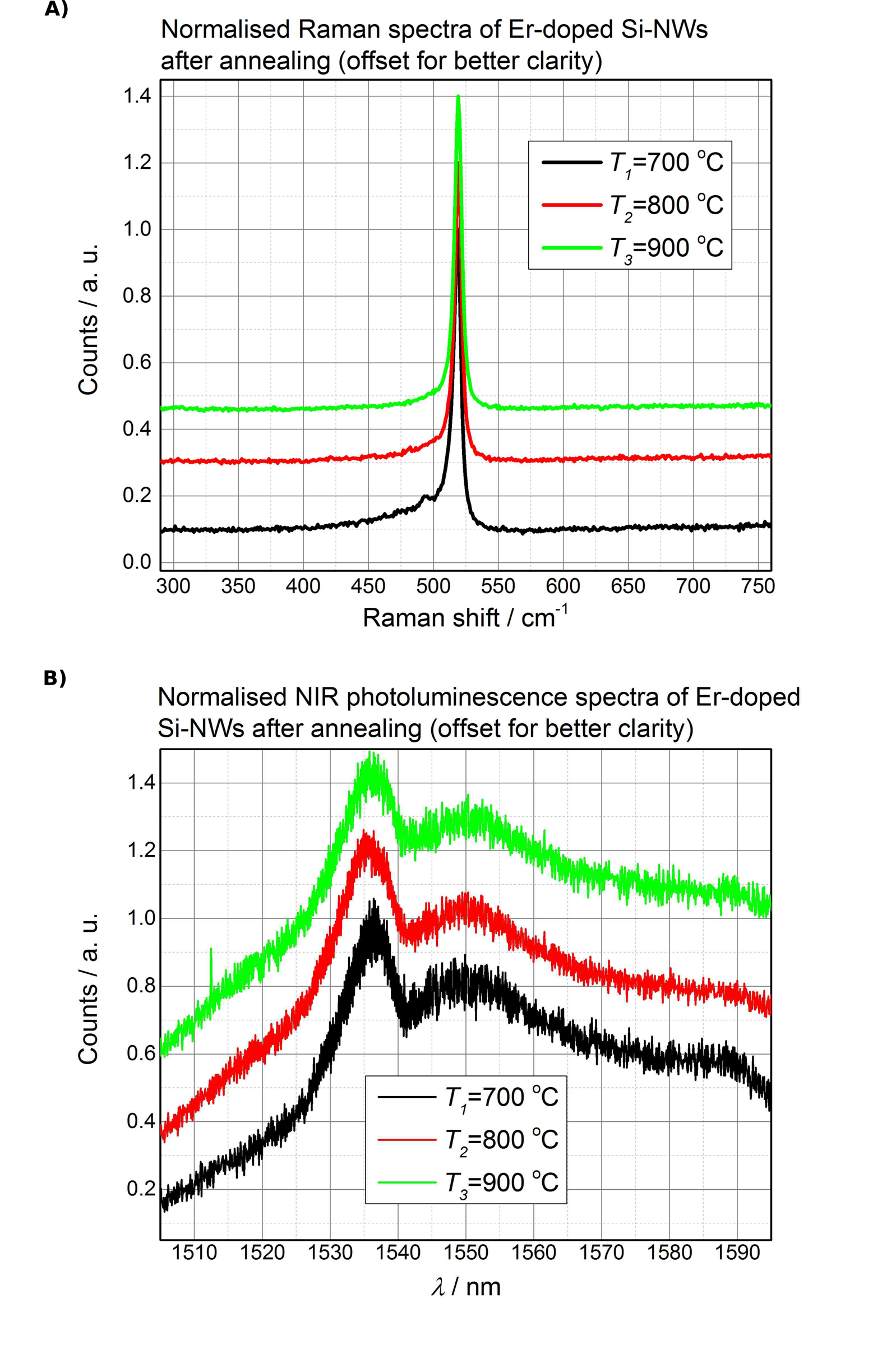} 
\caption[An example of a floating figure]{Spectra of silicon nanowires at room temperature after annealing for one hour at three different temperatures. A) Raman spectra around the first order optical phonon for silicon is shown. B) Photoluminescence spectra of the optical activated erbium atoms in the silicon nanowires. } 
\label{RSpectra} 
\end{figure}

\IEEEPARstart{T}{}he crystalline silicon nanowires are grown from gold catalyst particles via the vapour- liquid-solid process by us at Namlab in Dresden \cite{WMWeber}. The doping with Arsenic and Erbium was achieved using ion implantation at Australian National University (ANU) in Canberra. Afterwards, the catalyst particle is removed via the "Aqua Regia"-process before the crystal structure is restored in a furnace.  Fig. \ref{RSpectra} shows the spectra after annealing for $1$ hour at three different temperatures. In figure \ref{RSpectra} A) the silicon raman spectra peak at $520$ cm$^{-1}$ of the first order optical phonon 
is visible which proves the restored crystal structure. Figure \ref{RSpectra} B) shows the photoluminescence peak at $1550$ nm of the erbium transition from the first excited state to the ground state $^4$I$_{13/2}$ $\Longrightarrow$ $^4$I$_{15/2}$ which indicates the optical activation of erbium in the silicon crystal \cite{AJKenyon}. 
Finally, a high quality silicon dioxide is grown to passivate the nanowires \cite{AHeinzig}. 
\\
After this preparation, we deposit the nanowires on a thermally oxidised $4"$-silicon wafer with a $200$ nm SiO$_2$. The substrate of the wafer is highly doped to work as a global back gate which can control the carrier density in the nanowires. A positive resist is applied on the substrate surface and via electron beam lithography the mask is written. To enable proper contacts with the deposited metal a short HF dip will be done to remove the grown oxide from the nanowires. Nickel is used for deposition because it can be alloyed with silicon and creates a sharp Schottky barrier in the nanowire \cite{WMWeber}. The length of the silicidation segments can be controlled via time and temperature \cite{AHeinzig}.
\\
Once these fabrication steps are done the doped silicon nanowires can be characterised in our dilution refrigerator at temperatures down to $20$ mK. The experiments will include measurements using Lock-In technology at different magnetic fields (up to $9$ T), gate voltages and temperatures.

\section{Conclusion}
\IEEEPARstart{W}{}e believe due to this approach we have a huge versatility in fabricating different confined quantum systems in $1$- and $0$-dimensions and can contribute to a better understanding of those such as looking at spin and defect states in the nanowire and surroundings.

\section*{Acknowledgment}

\IEEEPARstart{T}{}his research was conducted by the Australian Research Council Centre of Excellence for Quantum Computation and Communication Technology (project number CE110001027). This work has been supported by the Melbourne International Fee Remission Scholarship (MIFRS) and the Melbourne International Research Scholarship (MIRS). The Department of Electronic Materials Engineering at the Australian National University is acknowledged for their support by providing access to ion implanting facilities. L. H. W. v. B. wishes to acknowledge the MMI seed funding grant.

%






\end{document}